\documentclass[11pt]{article}
\usepackage{graphicx}

\input{pubboard/babarsym}
\def\Dp      {\ensuremath{D^+}}
\def\Dm      {\ensuremath{D^-}}
\def\Dstar   {\ensuremath{D^{*}}}
\def\Dstarp  {\ensuremath{D^{*+}}}
\def\Dstarm  {\ensuremath{D^{*-}}}

\newcommand{\btoKzKz}{\ensuremath{\Bz\to \Kz\Kzb}}
\newcommand{\btoKsKs}{\ensuremath{\Bz\to \KS\KS}}

\newcommand{\etal}{{\it et al.}}

\def\fish    {\ensuremath{\cal F}}

\def\thsph    {\ensuremath{\theta_{\scriptscriptstyle S}}}

\newcommand{\BABARPubYear}    {01}

\newcommand{\BABARConfNumber} {04}
\newcommand{\SLACPubNumber} {8978}

\long\def\inst#1{\par\nobreak\kern 4pt\nobreak
    {\it #1}\par\vskip 10pt plus 3pt minus 3pt}

\begin{document}
{\pagestyle{empty}

\begin{flushright}
\babar-CONF-\BABARPubYear/\BABARConfNumber \\
SLAC-PUB-\SLACPubNumber \\
September, 2001 \\
\end{flushright}

\par\vskip 3cm

\begin{center}
{
\Large \bf
Search for \boldmath{$B$} Decays into $\Kz\Kzb$
}
\end{center}
\bigskip

\begin{center}
\large The \babar\ Collaboration\\
\mbox{ }\\
\today
\end{center}
\bigskip \bigskip

\begin{center}
\large \bf Abstract
\end{center}
We present preliminary results of a search for $\Bz\to\Kz\Kzb$ decays 
in the $\KS\KS$ final state using a sample of approximately $23$ 
million \BB\ pairs collected by the \babar\ detector at the \pep2\ 
asymmetric $B$ Factory at SLAC.  We find no evidence for a signal and 
set an upper limit on the branching fraction 
of $7.3\times 10^{-6}$
at the $90\%$ confidence level.  

\vfill
\begin{center}
Submitted to the 9$^{th}$ International Symposium 
on
Heavy Flavor Physics, \\
9/10---9/13/2001, Pasadena, CA, USA
\end{center}

\vspace{1.0cm}
\begin{center}
{\em Stanford Linear Accelerator Center, Stanford University, 
Stanford, CA 94309} \\ \vspace{0.1cm}\hrule\vspace{0.1cm}
Work supported in part by Department of Energy contract DE-AC03-76SF00515.
\end{center}

\newpage

\begin{center}
\small

The \babar\ Collaboration,
\bigskip

B.~Aubert,
D.~Boutigny,
J.-M.~Gaillard,
A.~Hicheur,
Y.~Karyotakis,
J.~P.~Lees,
P.~Robbe,
V.~Tisserand
\inst{Laboratoire de Physique des Particules, F-74941 Annecy-le-Vieux, France }
A.~Palano,
A.~Pompili
\inst{Universit\`a di Bari, Dipartimento di Fisica and INFN, I-70126 Bari, Italy }
% ...

\end{center}\newpage

\setcounter{footnote}{0}
\section {Introduction}

The study of $B$ meson decays into charmless hadronic final states plays
an important role in the understanding of \CP\ violation in the $B$ system.
Measurements of rates and \CP-violating asymmetries for $B$ decays into 
charmless two-body final states provide information on the angles $\alpha$ and 
$\gamma$ of the Unitarity Triangle.  However, in contrast to the 
theoretically clean determination of the angle $\beta$ in $B$ decays to charmonium 
final states~\cite{sin2beta}, the extraction of \CP-violation parameters in
charmless decays is complicated by hadronic uncertainties that are difficult
to calculate from first principles.  Accurate branching fraction measurements 
provide critical tests of theoretical models that are needed to obtain reliable 
information on $\alpha$ and $\gamma$.

The \babar\ collaboration has recently published \cite{ourPRL} measurements of
the branching fractions for $B$ meson decays to the charmless hadronic final 
states $\pip\pim$, $\Kp\pim$, $\Kp\piz$, $\Kz\pip$ and $\Kz\piz$, upper limits 
on the decays to $\pip\piz$ and $\Kz\Kp$ and the results of a search for charge 
asymmetries in the modes $\Bz\to \Kp\pim$, $\Bu\to \Kp\piz$ and 
$\Bu\to \Kz\pip$.\footnote{Charge conjugate modes are assumed throught this
paper.}  In this paper we report preliminary results of a search for
$\btoKzKz$ decays through detection of the $\KS\KS$ final state.  
Although the decay rate for \btoKzKz\ is expected to be small 
($10^{-6}$--$10^{-7}$) in the Standard Model~\cite{kkbarBR}, final state 
rescattering effects can lead to enhancement of the branching fraction and 
the possibility of large strong phases, with correspondingly large 
\CP-violating charge asymmetries~\cite{kkbarBR,kkbarfsi}.  Such rescattering 
effects may also have consequences for constraints on $\gamma$ derived from 
$\B\to K\pi$ decays~\cite{kpifsi}.  Observation of the $\Kz\Kzb$ decay mode 
would provide important information about the strength of final state 
rescattering in charmless $B$ decays.

\section {Data Sample}

The data used in this analysis were collected with the \babar\ detector
at the \pep2\ $\epem$ storage ring.  The sample corresponds to an 
integrated luminosity of $20.6\invfb$ taken near the
\FourS\ resonance (``on-resonance'') and $2.6\invfb$ taken at a 
center-of-mass (CM) energy $40\mev$ below the \FourS\ resonance
(``off-resonance''), which are used for continuum background studies.
The on-resonance sample corresponds to $(22.57\pm 0.36)\times 10^6$ 
\BB\ pairs.  The collider is operated with asymmetric beam energies, 
producing a boost ($\beta\gamma = 0.56$) of the \FourS\ along the 
collision axis.  The boost increases the momentum range of 
two-body $B$ decay products from a narrow distribution centered near 
$2.6\gevc$ to a broad distribution extending from $1.7$ to $4.3\gevc$.

The \babar\ detector is described in detail in Ref.~\cite{babarnim}.  
The primary detector element used in this analysis is the tracking system,
which consists of a $5$-layer, double-sided, silicon vertex detector and a 
$40$-layer drift chamber filled with a gas mixture of helium ($80\%$) 
and isobutane ($20\%$).  Both tracking detectors operate within a 
$1.5\,{\rm T}$ superconducting solenoidal magnet.  

\section {Event Selection and \boldmath{$\KS$} Reconstruction}

Hadronic events are selected based on track multiplicity and event topology.
Backgrounds from non-hadronic events are reduced by requiring the
ratio of Fox-Wolfram moments $H_2/H_0$ \cite{fox} to be less than $0.95$
and the sphericity \cite{spheric} of the event to be greater than $0.01$.

Candidate $\KS$ mesons  are reconstructed in the $\pip\pim$ final
state from pairs of oppositely charged tracks that form a well-measured 
vertex and have an invariant mass within $11.2\mevcc$ ($3.5\sigma$) of the 
nominal \KS\ mass~\cite{PDG}.  The measured proper decay time of the \KS\ 
candidate is required to exceed $5$ times its error.  Figure~\ref{ksmass} 
shows the invariant mass distribution of an inclusive sample of high
momentum ($>1\gevc$) \KS\ candidates.

\begin{figure}[!tbp]
\begin{center}
\begin{minipage}[h]{8.0cm}
\includegraphics[width=8.0cm]{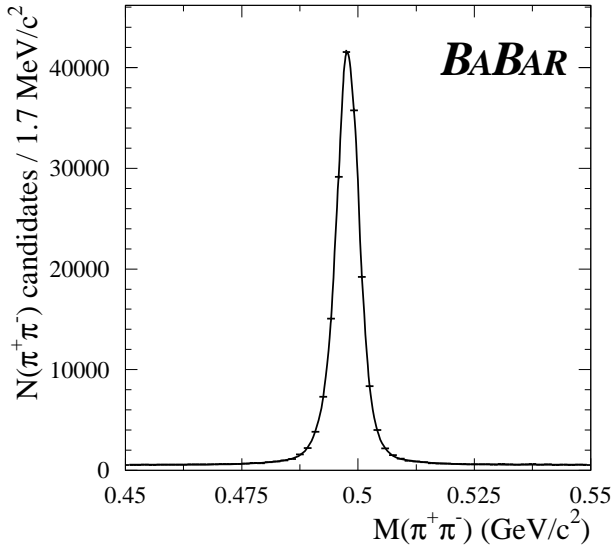}
\end{minipage}
\end{center}
\caption {Invariant mass of $\pip\pim$ pairs in the inclusive \KS\ sample
selected with requirements
on the angle between the flight direction and momentum of the \KS, and on
the transverse momenta of the decay products relative to the
\KS\ momentum.}
\label{ksmass}
\end{figure}

\section {\boldmath{$B$} Reconstruction and Background Rejection}
\label{brecosec}

$B$ meson candidates are reconstructed by combining pairs of
$\KS$ candidates.  The kinematic constraints provided by
the \FourS\ initial state and relatively precise knowledge of the
beam energies are exploited to efficiently identify $B$ candidates.  
We define a beam-energy substituted mass 
$\mes = \sqrt{E^2_{\rm b}-\mathbf{p}_B^2}$, 
where $E_{\rm b} =(s/2 + \mathbf{p}_i
\cdot\mathbf{p}_B)/E_i$, $\sqrt{s}$ and
$E_i$ are the total energies of the \epem\ system in the CM
and lab frames, respectively, and $\mathbf{p}_i$ and $\mathbf{p}_B$ are 
the momentum vectors in
the lab frame of the \epem\ system and the $B$ candidate, respectively.  
The \mes\ resolution for $B$ decays into all-charged final states is 
dominated by the beam energy spread and is determined
to be $2.6\pm 0.1 \mevcc$ from a Gaussian fit to a large sample of fully 
reconstructed $B$ decays.  Candidates are selected in the range 
$5.2<\mes<5.3\gevcc$.  A sideband region is defined as $5.2<\mes<5.26\gevcc$.

We define an additional kinematic parameter $\Delta E$ as the
difference between the energy of the $B$ candidate and half the energy
of the \epem\ system, computed in the CM system.  The $\Delta E$ 
distribution for signal events is a Gaussian centered near zero.  
The resolution on $\Delta E$ is estimated to be $21\pm 5\mev$
based on Monte Carlo simulated \btoKsKs\ decays and the observed
difference in widths between a control sample of fully reconstructed $B$ 
decays in data and
in Monte Carlo simulation.  We require $|\Delta E| < 0.1\gev$.
A sideband region is defined as $0.1 < |\Delta E| < 0.3\gev$.

Detailed Monte Carlo simulation, off-resonance data, and events in
on-resonance sideband regions are used to study backgrounds.  The 
contribution due to other $B$ meson decays, both from $b\to c$ and charmless 
decays, is found to be negligible from a detailed Monte Carlo study.  
The dominant source of background is 
from random combinations of true \KS\ mesons produced in the $\epem\to \qqbar$ 
continuum events (where $q=u$, $d$, $s$, or $c$).  
In the CM frame this background 
typically exhibits a two-jet structure, in contrast to the isotropic 
decay of $\BB$ pairs produced in \FourS decays.
We exploit the topology difference between signal and background by
making use of two event-shape quantities.  

The first variable is the angle 
$\thsph$ between the sphericity axes of the $B$ candidate and of the 
remaining tracks and photons in the event.  The distribution
of $|\cos\thsph|$ in the CM frame is strongly peaked near $1$ for
continuum events and is approximately uniform for \BB\ events.  We
require $|\cos\thsph| < 0.9$, which rejects $66\%$ of the background
remaining at this stage of the analysis.

The second quantity is a Fisher discriminant
${\cal F}$ constructed from the scalar sum of the CM momenta of
all tracks and photons (excluding the $B$ candidate decay products) flowing
into nine concentric cones centered on the thrust axis of the $B$
candidate.  Each cone subtends an angle of $10^\circ$
and is folded to combine the forward and backward intervals.  Monte
Carlo samples are used to obtain the values of the Fisher coefficients, which
are determined by maximizing the statistical separation between signal and
background events.  Figure~\ref{fig:fishersig} shows distributions of 
$\cal F$ for Monte Carlo simulated \btoKsKs\ decays and background events in 
the \mes\ sideband region. No cut is applyed on ${\cal F}$, instead the
distributions for signal and background events 
are included in a maximum likelihood as described in the next section.

A total of 286 candidates in the Run1 on-resonance data satisfy 
our selection criteria ( $|\cos\thsph| < 0.9$, $5.2<\mes<5.3\gevcc$ and 
$|\Delta E| < 0.1\gev$) and enter the maximum likelihood fit.
The total selection efficiency is $(36.6 \pm 4.6)\%$, where the error is 
dominated 
by uncertainty on the \KS\ reconstruction efficiency ($6\%$ relative 
error per \KS).  

\begin{figure}[!tbp]
\begin{center}
\begin{minipage}[h]{8.0cm}
\includegraphics[width=8.0cm]{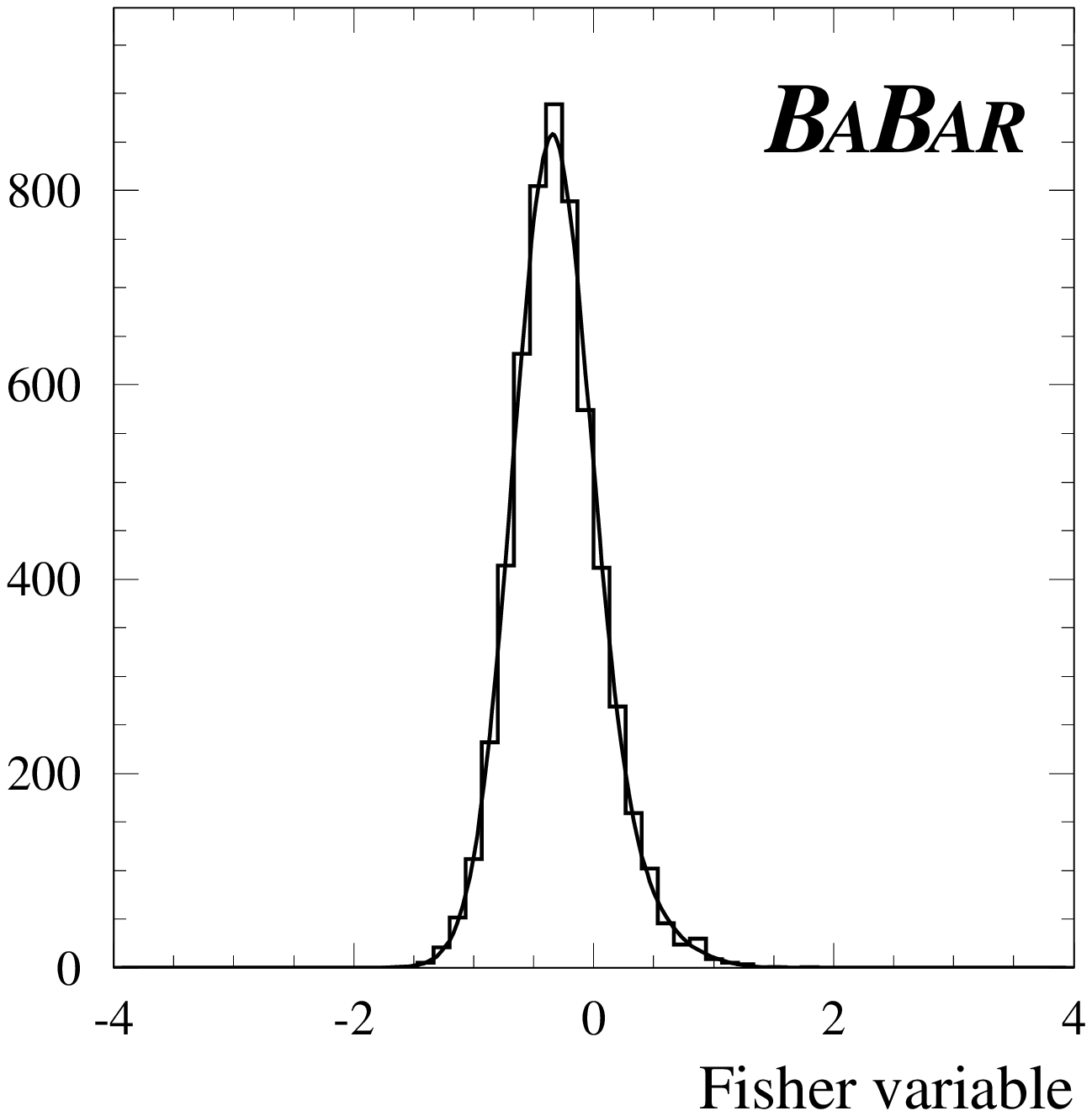}
\end{minipage}
\begin{minipage}[h]{8.0cm}
\includegraphics[width=8.0cm]{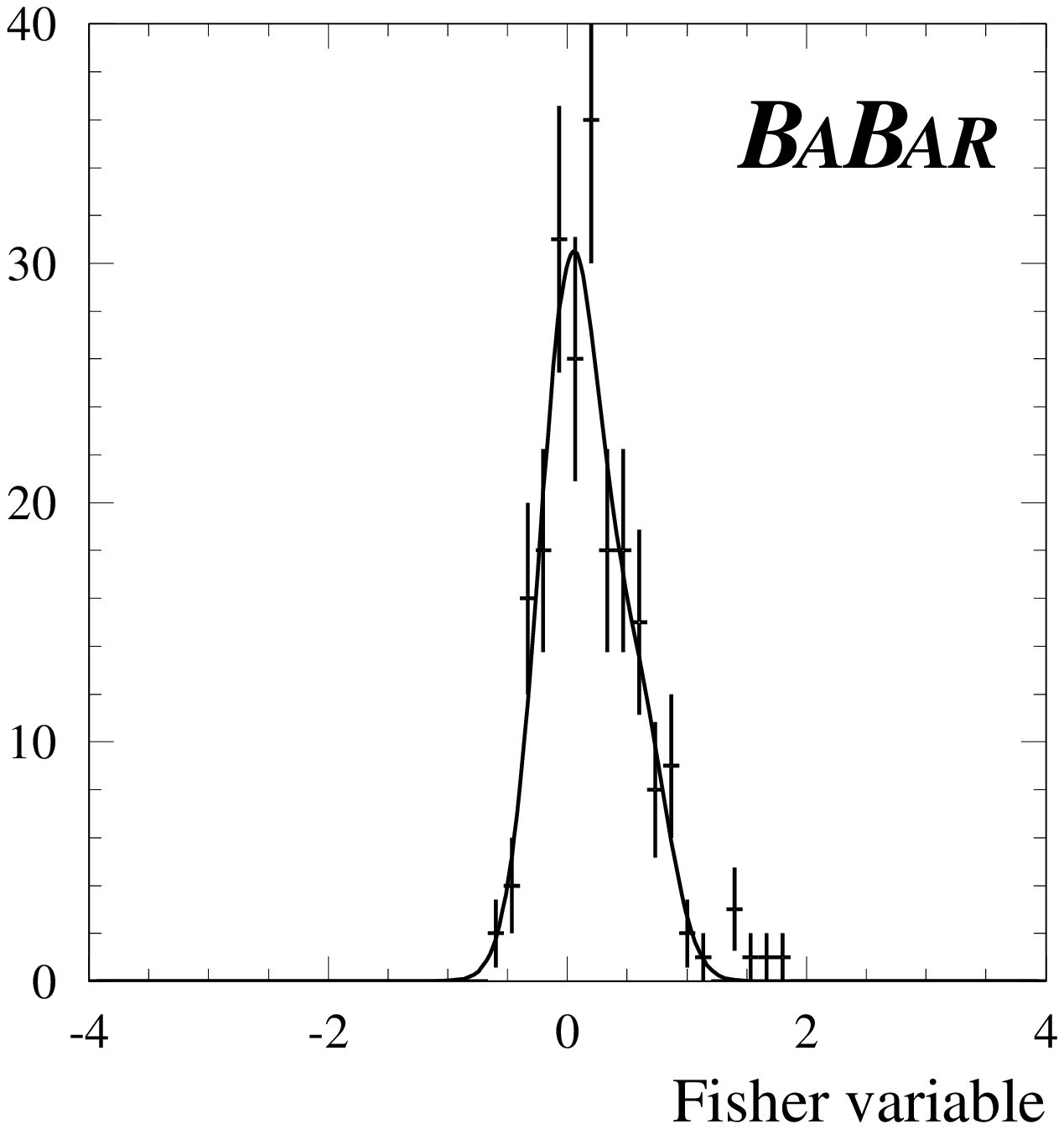}
\end{minipage}
\end{center}
\caption
{ Distributions of Fisher discriminant output for Monte Carlo simulated 
\btoKsKs\ decays (left) and events in the on-resonance \mes\ sideband data 
(right).}
\label{fig:fishersig}
\end{figure}

\section{Signal Extraction}

Signal and background yields are determined from an unbinned maximum 
likelihood fit using \mes, $\Delta E$ and \fish.  The likelihood is defined as
\begin{equation}
{\cal L} = e^{-(N_S+N_B)}\prod_i^N\left(N_S{\cal P}_i^S(\mes,\Delta E,{\fish})
+N_B{\cal P}_i^B(\mes, \Delta E,{\fish})\right),
\end{equation}
where $N_S$ and $N_B$ are the fitted number of signal and background
events, respectively; $N$ is the total number of events entering the
fit; and ${\cal P}_i$ is the product of probability density functions (PDFs) 
for \mes, $\Delta E$, and \fish~that are assumed to be uncorrelated.  
The quantity $-2\ln {\cal L}$ is minimized 
with respect to the fit parameters $N_S$ and $N_B$.

The parameters for background \mes\ and $\Delta E$ PDFs are determined from 
events in on-resonance $\Delta E$ and \mes\ sideband regions,
respectively.  The \mes\ shape is parameterized by a threshold 
function~\cite{argus} $f(\mes)\propto \mes\sqrt{1-x^2}\exp[-\xi(1-x^2)]$, where
$x=\mes/m_0$ and $m_0$ is the average CM beam energy.  The background shape in 
$\Delta E$ is parameterized as a second-order polynomial.
The signal distributions for \mes\ and $\Delta E$ are described by Gaussians,
where the \mes\ mean and width are determined from a sample of fully 
reconstructed $B$ decays while the $\Delta E$ mean and width are estimated 
from Monte Carlo simulated \btoKsKs\ decays and scaled according to 
the observed difference between a control sample of fully reconstructed $B$ 
decays in data and in Monte Carlo simulation. 
Events in on-resonance \mes\ sideband regions and Monte Carlo 
simulated signal decays are used to parameterize  as the sum of two Gaussians
the Fisher discriminant PDFs for background and signal.
Alternative parameterizations
for ${\cal F}$, obtained from off-resonance data (for background) and fully 
reconstructed $B$ decays (for signal), are used to determine systematic 
uncertainties.

The fitted number of signal events is $N_S = 3.4^{+3.4}_{-2.4}$, where the
error is statistical only.  Figure~\ref{fig:ksks_mes_ml} 
shows the \mes\ and  $\Delta E$ distributions for events satisfying the
selection criteria and additional requirements on the likelihood ratio
$N_S{\cal P}_S/(N_S{\cal P}_S+N_B{\cal P}_B)$, where the probabilities include
all variables ($\mes$, $\Delta E$, ${\cal F}$) except the one being plotted.  
The likelihood ratio cuts are chosen to minimize the upper limit on the 
branching fration, and the curves represent the fit 
result scaled by the efficiency of the additional requirements.  
The statistical significance of the signal yield is $1.5$, calculated as 
the square root of the change in $-2\ln{\cal L}$ when the yield is fixed to 
zero.  We conclude there is no evidence for a signal and calculate a $90\%$ 
confidence level Bayesian upper limit as 
the value of $N_{\rm UL}$ for which 
$\int_0^{N_{\rm UL}} {\cal L}_{\rm max}\,dN_S/\int_0^\infty 
{\cal L}_{\rm max}\,dN_S = 0.90$, where ${\cal L}_{\rm max}$ is 
the likelihood as a function of $N_S$, maximized with respect to the 
remaining fit parameter ($N_B$).  The resulting upper limit on the yield is
$N_{\rm UL} = 9$.
The fitting procedure has been validated with extensive Monte Carlo studies.

\begin{figure}[!tbp]
\begin{center}
\begin{minipage}[h]{8.0cm}
\includegraphics[width=8.0cm]{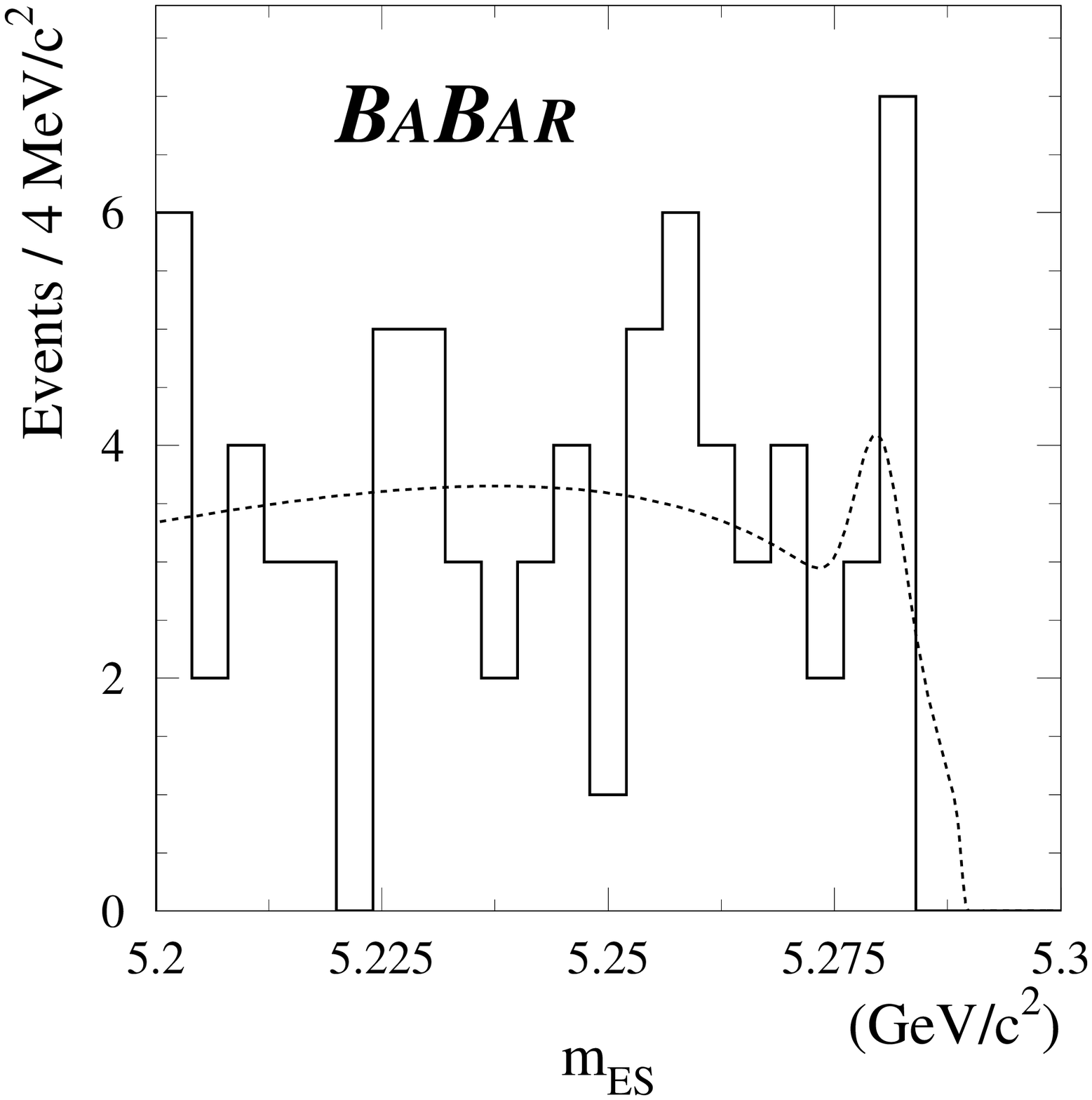}
\end{minipage}
\begin{minipage}[h]{8.0cm}
\includegraphics[width=8.0cm]{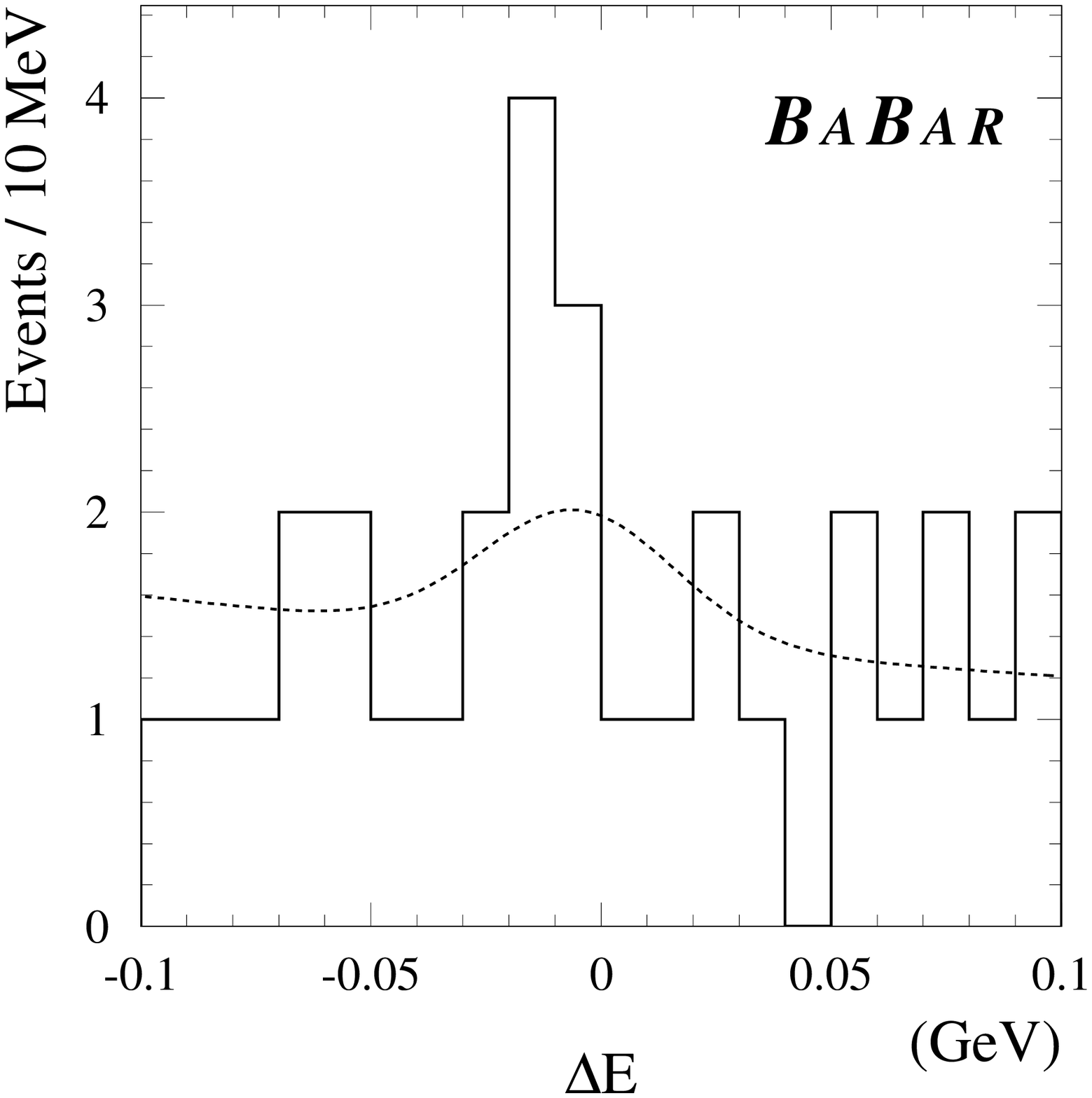}
\end{minipage}
\end{center}
\caption{Distributions of \mes\ (left) and $\Delta E$ (right) after
additional requirements on likelihood ratios.  The curves represent
projections of the maximum likelihood fit result.}
\label{fig:ksks_mes_ml}
\end{figure}

\section{Branching Fraction Results}
The branching fraction is defined as
\begin{equation}
\BR(\btoKzKz) = \frac{1}{\BR(\Kz\Kzb \to \KS\KS) \cdot 
\BR(\KS \to \pip\pim)^2}\frac{N_S}{\epsilon \cdot N_{\BB}},
\label{brksks}
\end{equation}
$\epsilon$ is the total $\KS\KS$ selection efficiency, 
$N_{\BB} = (22.57 \pm 0.36)\times 10^6$ is the total number of $\BB$ pairs 
in the dataset, and $\BR(\KS \to \pip\pim)= 0.6861$~\cite{PDG}.
We assume the Standard Model prediction that $\Bz\to\KS\KS$ proceeds 
through the $\Kz\Kzb$ intermediate state (as opposed to $\Kz\Kz$ or 
$\Kzb\Kzb$) and use $\BR(\Kz\Kzb\to\KS\KS) = 0.5$.\footnote{Assuming
conservation of angular momentum and $CPT$ invariance, the decay 
$\Bz\to\Kz\Kzb\to\KS\KL$ is forbidden.}  Implicit in 
Eq.~\ref{brksks} is the assumption of equal branching fractions for 
$\Y4S\to\Bz\Bzb$ and $\Y4S\to\Bu\Bub$.  

Systematic uncertainties on the branching fraction arise primarily from
uncertainty on the \KS\ reconstruction efficiency and uncertainty on $N_S$ 
due to imperfect knowledge of the PDF shapes.  The latter is estimated either 
by varying the PDF parameters within $1\sigma$ of their measured uncertainties 
or by substituting alternative PDFs from independent control samples.
Table~\ref{tab:syst} summarizes the various sources of systematic error on
$N_S$, where the total error is calculated as the sum in quadrature of the
individual uncertainties.

\begin{table}[!tbp]
\begin{center}
\caption {Summary of absolute systematic errors $(\sigma_{N_S})$ on the 
signal yield due to imperfect knowledge of the PDF shapes.  The total 
error is calculated as the sum in quadrature of the individual uncertainties.}
\begin{tabular}{ccc}
\hline\hline
Source & $\sigma_{N_S}$ \rule[-2mm]{0mm}{6mm} \\ \hline
\multicolumn{1}{l}{\boldmath $\mes$} \hspace{1.0cm} \rule[-1mm]{0mm}{5mm}\\
signal & $^{+0.3}_{-0.4}$ \\
background & $\pm 0.4$ \\
\multicolumn{1}{l}{\boldmath $\Delta E$} \hspace{1.0cm} \rule[-1mm]{0mm}{7mm}\\
signal & $^{+0.1}_{-0.4}$ \\
background & $-$ \\
\multicolumn{1}{l}{\boldmath ${\cal F}$} \hspace{1.0cm} \rule[-1mm]{0mm}{7mm}\\
signal & $\pm 3.2$ \\
background & $\pm 1.2$  \rule[-4mm]{0mm}{0mm} \\\hline
total  & $\pm 3.5$ \\
\hline\hline
\end{tabular}
\label{tab:syst}
\end{center}
\end{table}

We measure a central value branching fraction of 
$\BR(\btoKzKz) = (1.8^{+1.8}_{-1.2}\pm 1.8)\times 10^{-6}$, where
the first error is statistical and the second is systematic.
An upper limit is calculated by increasing $N_{\rm UL}$ and decreasing 
the efficiency by their respective systematic errors.  We find 
$\BR(\btoKzKz) < 7.3 \times 10^{-6}$ at the $90\%$ confidence level.
This result is a significant improvement over the existing upper limit 
from the CLEO Collaboration~\cite{cleokzkz}, and is approaching the upper
range of current theoretical estimates.

\section{Acknowledgements}
We are grateful for the 
extraordinary contributions of our \pep2\ colleagues in
achieving the excellent luminosity and machine conditions
that have made this work possible.
The collaborating institutions wish to thank 
SLAC for its support and the kind hospitality extended to them. 
This work is supported by the
US Department of Energy
and National Science Foundation, the
Natural Sciences and Engineering Research Council (Canada),
Institute of High Energy Physics (China), the
Commissariat \`a l'Energie Atomique and
Institut National de Physique Nucl\'eaire et de Physique des Particules
(France), the
Bundesministerium f\"ur Bildung und Forschung
(Germany), the
Istituto Nazionale di Fisica Nucleare (Italy),
the Research Council of Norway, the
Ministry of Science and Technology of the Russian Federation, and the
Particle Physics and Astronomy Research Council (United Kingdom). 
Individuals have received support from the Swiss 
National Science Foundation, the A. P. Sloan Foundation, 
the Research Corporation,
and the Alexander von Humboldt Foundation.



\begin{thebibliography}{99}

\bibitem{sin2beta}
\babar\ Collaboration, B.~Aubert {\it et al.}, \jprl{87}, 091801 (2001);
BELLE Collaboration, K.~Abe {\it et al.}, \jprl{87}, 091802 (2001);
ALEPH Collaboration, R.~Barate {\it et al.}, \plb{492}, 259 (2000);
CDF Collaboration, T.~Affolder {\it et al.}, \jprd{61}, 072005 (2000);
OPAL Collaboration, K.~Ackerstaff {\it et al.}, \epjc{5}, 379 (1998).

\bibitem{ourPRL}
\babar\ Collaboration, B.~Aubert \etal,  SLAC-PUB-8838, hep-ex/0105061,
to appear in \jprlBase

\bibitem{kkbarBR}
M.~Gronau and L.~Rosner, \jprd{58}, 113005 (1998);
C.H.~Chen and H-n~Li, \jprd{63}, 014003 (2000); 
D.~Du \etal, hep-ph/0108141.

\bibitem{kkbarfsi}
B.~Blok, M.~Gronau, and J.~Rosner, \jprl{78}, 3999 (1997);
A.J.~Buras, R.~Fleischer, and T.~Mannel, \npb{533}, 3 (1998);
D.~Atwood and A.~Soni, \plb{466}, 326 (1999);
P.~Zenczykowski, \jprd{63}, 014016 (2000).

\bibitem{kpifsi}
M.~Neubert, \plb{424}, 152 (1998);
D.~Atwood and A.~Soni, \jprd{58}, 036005 (1998);
A.F.~Falk, A.L.~Kagan, Y.~Nir, and A.A.~Petrov, \jprd{57}, 4290 (1998);
R.~Fleischer, \epjc{6}, 451 (1999);
K.~Agashe and N.G.~Deshpande, \plb{451}, 215 (1999);
I.~Caprini, L.~Micu, and C.~Bourrely, \jprd{60}, 074016 (1999);
M.~Gronau and D.~Pirjol, \jprd{61}, 013005 (1999).

\bibitem{babarnim}
\babar\ Collaboration, B.~Aubert \etal, SLAC-PUB-8569, to appear in
Nucl.\ Instrum.\ and Methods.


\bibitem{fox}
G.C.~Fox and S.~Wolfram, \jprl{41}, 1581 (1978).

\bibitem{spheric}
S.L.~Wu, Phys.\ Rep.\ {\bf 107}, 59 (1984).

\bibitem{PDG} Particle Data Group, D.E.~Groom \etal, Eur.\ Phys.\ J.\ 
C~{\bf 15}, 1 (2000).

\bibitem{argus}  
ARGUS Collaboration, H. Albrecht {\em et al.}, \jpl{B185}, 218 (1987).

\bibitem{cleokzkz} 
CLEO Collaboration, R. Godang \etal, \jprl{80}, 3456 (1997).

%
%
%
%
%
%
%
\end{thebibliography}
\end{document}